# Electric-field quenching of optically detected magnetic resonance in a π-conjugated polymer

Douglas L. Baird[a], Adnan Nahlawi[a], Kenneth Crossley[a], Kipp J. van Schooten[a], Mandefro Y. Teferi[a], Henna Popli[a], Gajadhar Joshi[a], Shirin Jamali[a], Hans Malissa[a], John M. Lupton[a,b], Christoph Boehme[a]

[a]Department of Physics and Astronomy, University of Utah, UT, USA

[b]Institut für Experimentelle und Angewandte Physik, Universität Regensburg, Germany

Electric fields are central to the operation of optoelectronic devices based on conjugated polymers since they drive the recombination of electrons and holes to excitons in organic light-emitting diodes but are also responsible for the dissociation of excitons in solar cells. One way to track the microscopic effect of electric fields on charge carriers formed under illumination of a polymer film is to exploit the fluorescence arising from delayed recombination of carrier pairs, a process which is fundamentally spin dependent. Such spin-dependent recombination can be probed directly in fluorescence, by optically detected magnetic resonance (ODMR). Depending on the relative orientation, an electric field may either dissociate or stabilize an electron-hole carrier pair. We find that the ODMR signal in a polymer film is quenched in an electric field, but that, at fields exceeding 1 MV/cm, this quenching saturates. This finding contrasts the complete ODMR suppression that was previously observed in polymeric photodiodes, indicating that exciton-charge interactions–analogous to Auger recombination in crystalline semiconductors–may constitute the dominant carrier-pair dissociation process in organic electronics.

## 1. Introduction

One of the longest-standing puzzles in photophysical phenomena in π-conjugated polymers is the process of charge generation following optical excitation [1]. Photoexcitation typically leads to the population of a higher-lying intramolecular excited state of the extended π-electron system, a state which can be regarded as an electrostatically bound electron-hole pair, i.e. an exciton. This pair can swiftly relax to the lowest excited state of the molecule–separated energetically from the ground state by the optical gap energy–by dissipating excess energy through molecular vibrations. But how does this exciton ultimately split to form free charge carriers, for example in a photodiode or a solar cell? Usually, such a splitting is facilitated by incorporating a potential gradient, which can be achieved by blending suitable materials together in a bulk heterojunction. Alternatively, one can envision splitting the exciton by an electric field [2], an effect which is easily demonstrated by the photoluminescence (PL) quenching of conjugated polymers [3-18].

Large electric fields, often on the order of MV/cm, play a central role in the operation of organic light-emitting diodes (OLEDs), where electrons and holes are injected from electrodes. In this case, the carriers can actually capture one another through their mutual electrostatic interaction to form intramolecular excitons, which ultimately give rise to luminescence. In the early years of research into polymer-based optoelectronics, it was not clear whether conjugated polymers behave more like one-dimensional semiconductors with strong dielectric screening, or molecular crystals with weak screening [19]. Conclusive evidence for the molecular-exciton picture of conjugated polymers ultimately derived from the response of the fluorescence of polymer films to electric fields, which demonstrates that the luminescence can be strongly quenched [3, 10]. Given such field-induced quenching, it may seem surprising that fields are also present in the electroluminescence process, where quenching evidently does not dominate. Indeed, even the extreme electric fields generated by a biased scanning-tunnelling microscope tip have proven insufficient to quench excitonic species, giving rise to tip-induced electroluminescence [20].

There are several ways to examine the nature of photogeneration of charge carriers in conjugated polymer films: far-infrared absorption features due to the sub-gap transitions that emerge from the dipole associated with the open shell [21]; direct dielectric relaxation measurements in the THz regime combined with photoexcitation [22]; direct photocurrent measurements, which often require an external bias and are not always free of artefacts [1, 23]; and delayed luminescence through geminate and non-geminate recombination of carriers [24-26]. Ultimately, the most direct way to probe the presence of a charge is by identifying its spin signature in magnetic resonance experiments [27-36]. This approach to conjugated polymers was recognized early on but has only received a limited amount of attention over the past 30 years. Charge carrier photogeneration can be followed by photocurrent-detected magnetic resonance or by a modulation of the non-geminate delayed luminescence contribution [37, 38], by flipping spin species in optically detected magnetic resonance (ODMR). In ODMR experiments, typically, electron and hole spin permutation symmetries are inverted, which changes singlets into triplets and vice versa [38]. Since the two carrier-pair species result in excitonic states with very different radiative recombination yields, changes to the spin states of the excitonic charge carrier precursor pair ensemble immediately impact the PL quantum yield. Alternatively, ODMR may be performed on spin-1 species, i.e. the triplet excited states [33], a technique which is, in principle, sensitive down to the single-molecule level [39, 40]. While there are many examples of electroluminescence-based ODMR in devices [27, 35, 41], there is only one report thus far on PL-based ODMR in an OLED structure [42].

The result of an external electric field on the PL-based ODMR signal of a conjugated polymer film is not immediately apparent: on the one hand, one may expect the field to dissociate charge carrier pairs, decreasing the precursor pair population, thereby quenching the ODMR signal. On the other hand, the field itself may quite obviously increase carrier generation, as is evidenced by increased photoconductivity [25], increasing the precursor pair population, in which case the ODMR signal should increase. The obvious way of testing these hypotheses is to perform ODMR on a working OLED structure in reverse bias. Surprisingly, this seemingly simple experiment was only reported very recently. Kanemoto *et al.* [42] demonstrated that the ODMR signal is indeed quenched under reverse biasing of an OLED, and at bias voltages substantially below those for which a reduction in PL intensity is observed due to field-assisted exciton dissociation. The quenching of ODMR was found to correlate with an increase in photocurrent of the OLED structure, implying that charge carriers are indeed removed from the device so that they are no longer available for magnetic-resonance controlled recombination. However, in a conventional OLED structure, the interaction with charge carriers, either injected under forward bias or formed due to light absorption under illumination in reverse bias, provide an additional quenching pathway for both excitons–which are responsible for PL–and exciton precursor states, i.e. the weakly coupled electron-hole pairs. As such, the results presented in Ref. 42 motivate the question whether it is current quenching or field dissociation which is responsible for the reduction in the ODMR signal. We therefore aim to test the effect of electric fields on the ODMR signal in the absence of photocurrents.

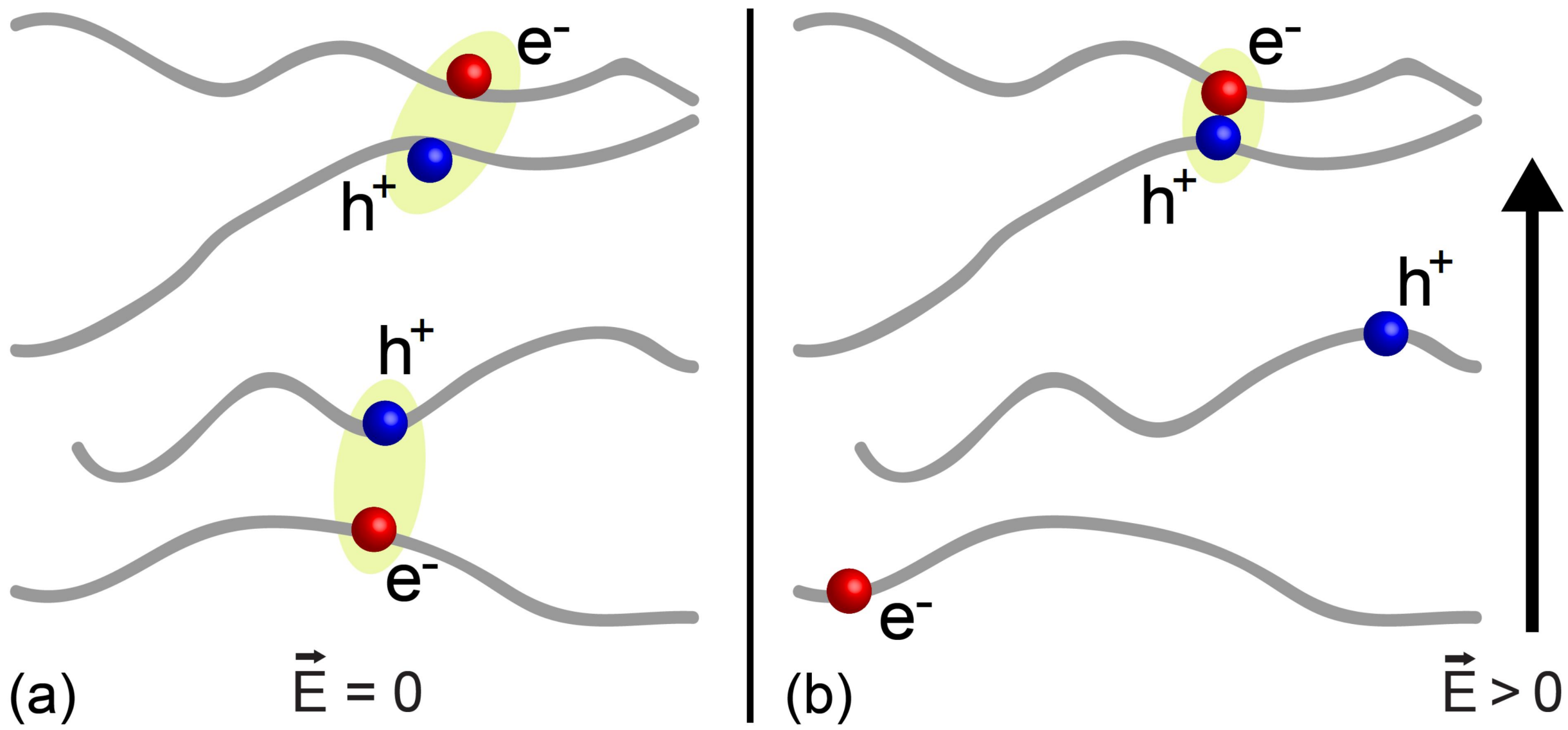

*Fig. 1. Effect of an external electric field on an electron-hole pair in a film of a conjugated polymer (grey). (a) An electron and hole can become Coulombically bound when situated on different polymer chains. Recombination does not occur until electron and hole reside on the same conjugated unit of the polymer. (b) Depending on orientation, an electric field will either push the carriers of the pair closer together or dissociate them. In the latter case, the Coulombic correlation between the carriers can break down and they turn into free charge carriers, so that recombination and the associated formation of a luminescent exciton become suppressed.*

Figure 1 summarises the hypothesis for the effect of an external electric field on the ODMR signal amplitude. At zero external field, inter-chain electron-hole pairs exist with different spatial separations and orientations, due to different chain conformations in the solid state [cf. panel (a)]. An external electric field may either push these carriers together, potentially promoting recombination, or drive their separation, as indicated in panel (b) [26]. In either case, a reduction in the overall ODMR amplitude is to be expected. However, in the presence of an electric field, PL quenching typically results from the spatial dissociation of electrons and holes within an exciton. This dissociation is an intermolecular effect and only occurs in films, and not in single polymer chains [43]. The process of PL quenching and the accompanying dissociation can be observed clearly in time-resolved experiments, where the delayed luminescence is detected following a voltage pulse [24, 26]. Such stabilization of charge-carrier pairs in an electric field can serve as a means of storing and temporally gating excitation energy in a polymer film [26]. It is therefore also conceivable that the ODMR signal rises under strong bias voltages.

## 2. Methods

The ODMR signal in conjugated polymers at room temperature usually originates from delayed recombination of short-lived intermediate electron-hole pairs which exist either in the singlet or the triplet state [37]. These pairs may recombine to singlet or triplet excitons, which makes charge carrier recombination fundamentally spin dependent. Under magnetic resonant excitation, the singlet and triplet pair populations are inverted, an effect which can be detected as a change in the fluorescence intensity [37]. A challenge in any magnetic resonance experiment involving electrical leads is to avoid the deterioration of the electromagnetic fields due to the interconnects. This is best achieved by using thin-film wiring based on indium tin oxide (ITO) or evaporated metal layers [44]. In addition, the fluorescence of the electrically contacted polymer film needs to be excited by a laser and detected by a photodiode. We found that, given the spatial constraints, commercial cavities for electron-paramagnetic resonance (EPR) are only poorly suited for this challenge. Limitations arise because of the combined challenges of finding a field-effect device scheme which does not significantly distort the resonator eigenmodes when a device is inserted while simultaneously allowing optical access with sufficiently high PL detection efficiency. We therefore designed a magnetic resonance probe-head specifically for the purpose of performing ODMR on organic field-effect device structures, working at sufficiently low magnetic fields to allow unfettered, free-space PL excitation and detection. Figure 2(a) sketches the experimental setup, and the device structure is shown in panel (b). The device is based on a conventional OLED layout as described in Ref. 44, with the crucial difference being that insulating layers are incorporated after each electrode to prevent charge injection and extraction by the electric field. Such insulation is best achieved with spin-on glass (SOG, Futurrex, Inc.) which can be deposited by spin coating to yield transparent, stable, planar films of thickness 200-500 nm. A first layer of SOG was deposited on a prepatterned ITO strip on a glass substrate, purchased from SPI Supplies (Structure Probe, Inc.). After drying and annealing of the SOG for 10 minutes at 100ºC, a layer of the conjugated polymer poly[2-methoxy-5-(2'-ethylhexyloxy)-p-phenylene vinylene] (MEH-PPV, purchased from American Dye Source) was spin-coated on top of the SOG layer from a toluene solution. The thickness of this active layer typically was on the order of 100 nm, and after annealing for 10 min at 100ºC, the structure was capped by a second SOG layer in order to provide electrical isolation from the top electrode, a thermally deposited 150-nm-thick layer of aluminium. We note that the detectability of PL ODMR signals in MEH-PPV films strongly depends on the film deposition parameters. In preparation for the experiments presented in the following, we studied the ODMR responses for a variety of deposition parameters, including deposition methods, spin frequencies, concentration of the toluene solutions, and annealing temperatures. We found that for the device structures discussed here, a significant ODMR response was consistently seen for MEH-PPV films deposited at room temperature from toluene solution with 10 g/l polymer concentration by spin casting with rotation frequencies of 1000 to 3000 rpm. We prepared several of the devices with epoxy encapsulation and several without. In both cases, only small photodegradation effects of the ODMR signals were seen.

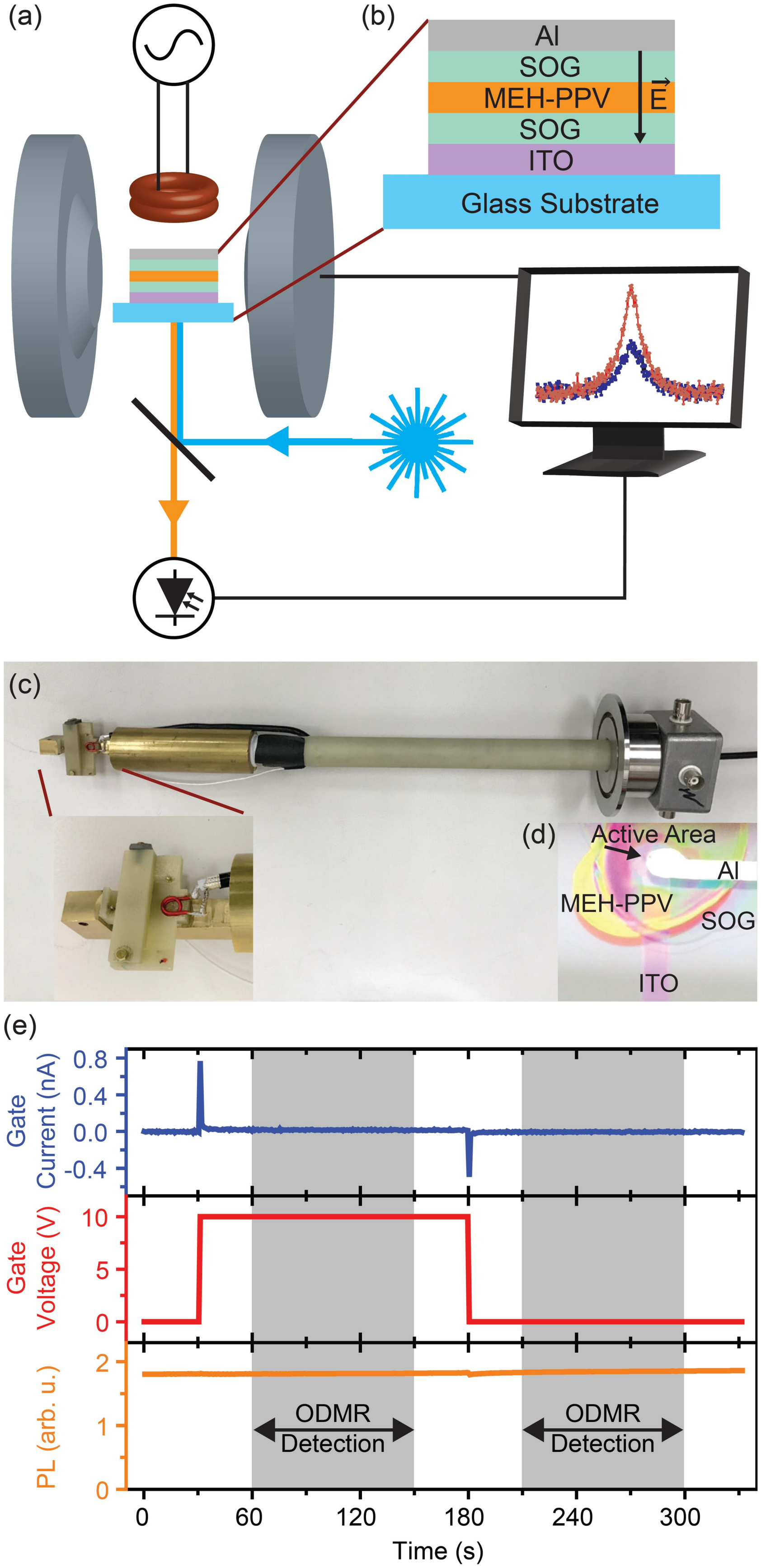
(a)
(b)
Al
SOG
MEH-PPV
SOG
ITO
Glass Substrate
$\vec{E}$
(c)
(d)
Active Area
Al
MEH-PPV
SOG
ITO
(e)
Gate Current (nA)
0.8
0.4
0.0
-0.4
Gate Voltage (V)
10
5
0
PL (arb. u.)
2
1
0
ODMR Detection
ODMR Detection
0
60
120
180
240
300
Time (s)

*Fig. 2. Measurement principle of electric-field-modulated photoluminescence (PL) optically detected magnetic resonance (ODMR) of a conjugated polymer film. (a) The device structure [cf. panel (b)] is placed between two coils of a magnet and radiofrequency (RF) radiation is supplied from a small single-loop coil. PL is excited in the device structure by irradiation with a laser beam with a wavelength of 405 nm (shown in blue), and the fluorescence (shown in orange) is collected using a photodiode after passing through a dichroic mirror. The sketches on the computer screen illustrate examples of lock-in detected ODMR spectra, plotted as differential PL intensity versus static magnetic field strength. (b) The capacitive device structure, which allows application of large electric fields without generating a photocurrent due to the photoexcitation, consists of a layer of indium-tin oxide (ITO), insulating layers of spin-on glass (SOG), and the active conjugated polymer material MEH-PPV. (c) Photograph of the probe-head with RF leads, and a close-up view of the sample holder with the capacitive device structure and the RF coil mounted. (d) Photograph of the capacitor structure. To minimize inductive effects and distortion of the RF field, thin-film Al and ITO wiring is used to contact the pixel. These wires are perpendicular to each other in order to minimize capacitance and leakage effects. The different colours of the device structure arise from optical interference effects. (e) Transients of device current and PL intensity as the voltage is turned on and off for one particular device. The brief current peaks coinciding with the rising and falling edges of the voltage pulse show displacement currents as the capacitor charges and discharges upon turning the voltage on and off, but the PL intensity is not significantly affected by the electric field. ODMR spectra are recorded both with the electric field turned on and off (shaded areas).*

The fabricated device structures were placed between the two coils of the iron-core magnet of a Bruker E580 EPR spectrometer to induce Zeeman-splitting of the electron and hole spin states. A radiofrequency (RF) field was applied using a single-turn coil oriented orthogonal to the magnet. Such a single-turn RF coil creates highly inhomogeneous and anisotropic RF field amplitudes, i.e. $B_1$ field strengths, but given the thickness of the photoactive polymer layer–on the order of 100 nm–such inhomogeneity only leads to small distributions of $B_1$ within the volume of the samples studied. Moreover, for the continuous-wave (c.w.) ODMR experiments reported in the following, weak magnitudes of $B_1$ are employed such that the spin resonantly induced singlet-to-triplet transition rates are small compared to both electronic transition rates as well as spin relaxation rates. Under such conditions, ODMR signal magnitudes are linear with $B_1$ [41] and therefore distributions of $B_1$ become irrelevant for the observed ODMR line shapes. The PL employed for ODMR detection was excited by a c.w. laser diode at a wavelength of 405 nm and a typical power of 15 mW, focused to a 1 mm diameter spot size on the sample, and detected by a silicon photodiode in series with a Femto variable gain high speed current amplifier (Model DHPCA-100). As sketched in Figure 2a, the blue excitation light and the resulting orange fluorescence of the polymer film are separated by a dichroic mirror in the beam path.

For the ODMR experiments discussed here, the change of PL intensity under RF excitation by an Agilent N5161A MXG signal generator is recorded as a function of Zeeman splitting, i.e. of static magnetic field strength $B_0$, using the internal lock-in detector of a Bruker E580 spectrometer in connection with rectangular amplitude modulation of the RF excitation at a frequency of 10 kHz. This gives rise to charge-carrier ODMR spectra as plotted in Fig. 2(a). Fig. 2(c) shows a photograph of the probe-head which is placed inside the magnet and contains both the RF leads and the connections for the static electric field. A close-up shows the device placed within the fiberglass (Garolite G10) holder designed to lie in the plane of the external magnetic field, with the RF coil also visible close to the sample. A photograph of the device structure is given in Fig. 2(d), revealing differently coloured regions due to interference effects in the dielectric stack. Crucially, we found it necessary to place the two thin-film electrodes orthogonally to each other to minimize leakage currents at the high electric fields applied. The active area over which the electric field is applied corresponds to a circular pixel of area 2 $mm^2$. Placing the leads orthogonally to each other also minimizes the capacitance of the structure, which was typically found to be on the order of 30 pF. We note that there are several other conceivable ways to apply the RF field, for example using a microwire integrated monolithically into the device structure [45], or a coplanar waveguide placed beneath the device [46, 47]. However, we found that the coplanar waveguide structures tended to become short circuited in the large area sample geometry optimized for PL detection, preventing magnetic resonance excitation.

Fig. 2(e) shows the basic operation of the capacitor device structure and the measurement cycle. A typical measurement consists of c.w. illumination of the polymer film with alternating application of the electric field. ODMR spectra are recorded by magnetic field sweeps over a duration of 90 s, both with an electric field and without. The voltage of up to 150 V was applied with a Keithley 2400 source-measure unit, which also allowed measurement of the device current. Crucially, due to the insulating device structure, the overall current was limited to small leakage currents which are below 100 nA for all devices even at highest biases, and below 30 nA at biases below 100 V, corresponding to maximal current densities of <5 $\mu A/cm^2$. The change of the leakage current with illumination was negligible, i.e. there was no detectable photocurrent, indicating that the gate resistance of these devices determined the leakage currents both in the presence and in absence of the illumination. During the rising and falling edge of the electric field pulse, a significant transient displacement current is observed, which is not linked to illumination but to the displacement current of the capacitor structure.

In capacitive geometries, electric fields can lead to PL quenching by electrostatically dissociating the electron and hole within the fluorescent exciton [26]. However, this effect is generally inhibited at high excitation fluences, both due to the build-up of space charge [48] and because of non-linear exciton-exciton interaction effects such as singlet-singlet annihilation [49], which promotes the build-up of depolarisation fields. Since shot noise limits the differential measurement of PL intensity in ODMR, which typically records changes on the order of 10 ppm, we can only perform PL measurements at high laser excitation intensities. As shown in the lower panel of Fig. 2(e), the PL intensity does not change appreciably during application of the electric field.

## 3. Results and discussion

The left panel of Fig. 3(a) shows an ODMR spectrum of an MEH-PPV film in a capacitor structure with no electric field applied. When an RF excitation with a frequency of 112 MHz is applied to the sample, a pronounced resonance feature is observed at a static magnetic field of 4.7 mT. The feature has an overall amplitude on the order of 30 ppm. We have previously demonstrated that the ODMR feature of MEH-PPV in PL originates from an electron-hole pair process [38,41]. This fact is apparent from the evolution of the coherent spin nutation in pulsed magnetic resonance experiments with driving pulse length: at low driving intensities either electron or hole are in resonance, whereas at high driving powers both precess together, giving rise to a doubling of the Rabi frequency in a spin-beating effect [38]. Since the critical driving power for such beating to occur is determined by the magnetic disorder experienced by the resonant spin species which arises primarily from hyperfine interactions, deuteration of the material offers a facile way to confirm the microscopic pair origin of spin beating in MEH-PPV ODMR [38]. In addition, the measured Rabi frequency provides an absolute metric of the spin of the resonant species, which is found to be s=½ [50]. What is not immediately apparent is how these spins are actually generated in the film, given the fact that photoabsorption primarily leads to the formation of tightly bound exciton species. It is most likely that a small fraction of excitons dissociate on chemical or structural defects within the material [51]. Much of this dissociation gives rise to geminate-pair formation, which retains the overall spin of the primary photoexcitation during recombination in delayed luminescence [26]. However, some carrier pairs may also separate completely to give rise to the intrinsic photoconductivity characteristic of even pristine materials [1].

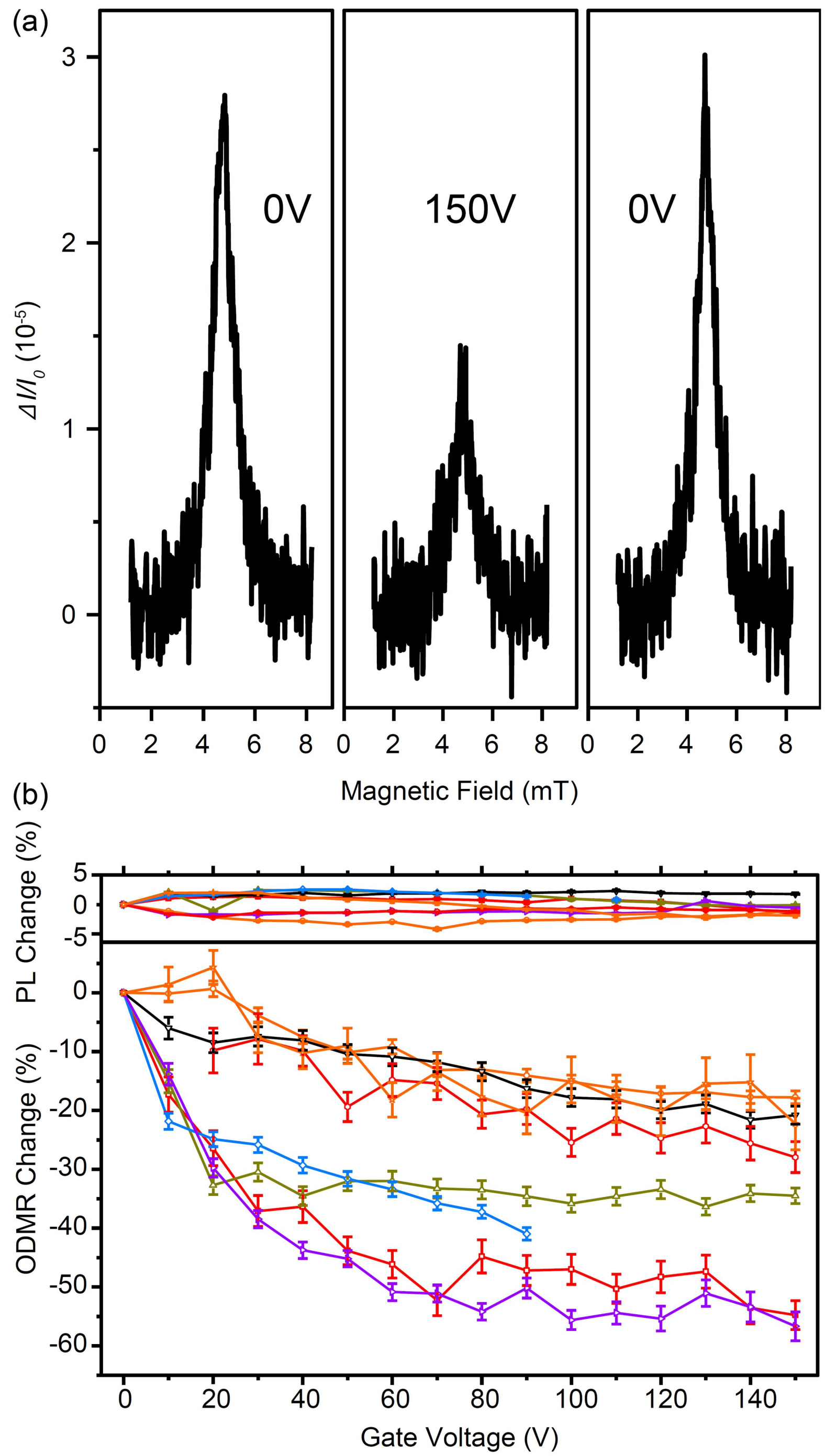

*Fig. 3. Quenching of the PL ODMR in an electric field. (a) Example of reversible ODMR quenching on a single device. (b) Bias dependence of ODMR quenching for 8 different devices together with the associated change in PL intensity. The ODMR quenching is significant, and appears to saturate at larger biases, while changes in the PL intensity are insignificant and do not correlate with the applied bias.*

Electric-field modulated ODMR reveals that, under application of a bias voltage of 150 V, corresponding to an electric field of approximately 1 MV/cm, the ODMR amplitude is almost halved, as shown in Fig. 3(a). The ODMR amplitude returns to its original value after removal of the electric field. From this simple observation two important conclusions can be drawn: under application of an electric field, the population of resonant electron-hole pair species is not increased. Conversely, considering prior PL-quenching experiments at high electric fields, one may have actually expected an enhancement. The fact that the ODMR signal does not increase under an electric field may, however, simply be related to the fact that no PL quenching is observed in these experiments at high fluences: a lack of PL quenching implies no formation of additional carrier pairs available for the resonance. Instead, the reduction of the ODMR signal under an external electric field implies the situation sketched in Fig. 1. Either carrier pairs are moved closer together, possibly recombining, and therefore become unavailable for manipulation by magnetic resonance; alternatively, they are pulled apart further by the electric field, thus entirely blocking recombination and therefore suppressing ODMR. Since the distribution of carrier pairs in the MEH-PPV film is, to a first approximation, expected to be isotropic, one would not anticipate that an external electric field be able to remove all carrier pairs for ODMR. This is analogous to carrier-pair-based delayed fluorescence in conjugated polymers, where an electric field pulse leads to both initial enhancement of delayed PL and subsequent suppression [26]. This simple intuitive argument is, however, evidently at variance with the recent report of complete ODMR quenching in OLED devices without insulating layers [42], suggesting that photocurrent flow can have an additional dramatic impact on the resonant pair population.

We tested 8 different devices over a broad field range, as summarized in Fig. 3(b). While all devices showed a certain degree of ODMR quenching at all applied electric fields, they also exhibit saturation of the quenching magnitude, typically on the order of 50%, at elevated fields. Small variations of at most a few percent are observed in the PL intensity, but these may be either positive or negative. Such a variation of the response of the PL intensity to electric fields has been reported previously and has been linked to the change in balance between prompt and delayed PL [52], carrier-assisted quenching of singlet excitons, and the build-up of space charge. However, in no case is an increase in the ODMR signal amplitude observed under application of an electric field. Similarly, the magnitude of the ODMR signal change is significantly larger than the observed PL changes in all cases which confirms that the observed changes of the ODMR signals are not caused by the changes of the PL intensities.

## 4. Conclusions

PL-based ODMR is a sensitive technique which can be applied to study various materials, including single molecules [39, 40]. Surprisingly, even though the technique has been used to gain insight into spin-dependent recombination processes in materials used for organic electronics [33], it has only very recently been applied to actual OLED devices [42]. In this previous report, complete suppression of ODMR from an OLED was observed under reverse biasing, and this suppression was attributed to assisted escape of all carriers out of the constituent pairs' mutual Coulomb well [42]. We find that complete suppression of ODMR is not observed when the photocurrent is inhibited in a capacitor structure with insulating layers. We conclude from this observation that electron-hole carrier pairs, which form the precursor to radiative excitons in OLEDs, are readily quenched by free mobile charges–probably more so than the tightly bound excitons. This conclusion is in agreement with the suggestion that the strong saturation of light intensity in triplet-exciton-based phosphorescent OLEDs is more closely related to charge-induced dissociation of the exciton precursor pairs rather than triplet-triplet annihilation between excitons [53]. Once photocurrent is eliminated, a universal saturation of the quenching of the ODMR signal at a finite value is observed, which strongly implies an isotropic distribution of charge-carrier pairs within the conjugated polymer film. We hypothesize that the quenching of the carrier-pair population by charges within an OLED device where the photocurrent is not inhibited [42] is analogous to Auger recombination in inorganic semiconductor crystals [54] and that such a process may therefore play a greater role in conjugated-polymer-based devices such as OLEDs and solar cells than previously anticipated.

**Acknowledgements**

We are indebted to Dr. Su Liu, Nick Borys, Marzieh Kavand, and Dr. Dongbo Li for helpful discussions regarding the PL quenching structures and to Andreas Sperlich for suggestions for improving the PL detection. This work was supported by the US Department of Energy, Office of Basic Energy Sciences, Division of Materials Sciences and Engineering under Award #DE-SC0000909. Kenneth Crossley was supported through the Utah MRSEC REU program (NSF project #DMR-1121252).